\documentclass[conference]{IEEEtran}
\IEEEoverridecommandlockouts
\usepackage{cite}
\usepackage{amsmath,amssymb,amsfonts,bm}
\usepackage{algorithmic}
\usepackage{graphicx}
\usepackage{textcomp}
\usepackage{float}
\usepackage{dblfloatfix}

\def\BibTeX{{\rm B\kern-.05em{\sc i\kern-.025em b}\kern-.08em
    T\kern-.1667em\lower.7ex\hbox{E}\kern-.125emX}}

\begin{document}
\title{Improved Simultaneous Multi-Slice Functional MRI Using Self-supervised Deep Learning
}

\author{\IEEEauthorblockN{Omer Burak Demirel$^{1,2}$,
        Burhaneddin Yaman$^{1,2}$,
        Logan Dowdle$^{2,3}$,
        Steen Moeller$^{2}$,
        Luca Vizioli$^{2,3}$,\\
        Essa Yacoub$^{2}$,
        John Strupp$^{2}$,
        Cheryl A. Olman$^{2}$,
        K\^{a}mil U\u{g}urbil$^{2}$
        and Mehmet Ak\c{c}akaya$^{1,2}$}
		\IEEEauthorblockA{
			$^{1}$Department of Electrical and Computer Engineering, University of Minnesota, Minneapolis, MN\\
			$^{2}$Center for Magnetic Resonance Research, University of Minnesota, Minneapolis, MN\\
			$^{3}$Department of Neurosurgery, University of Minnesota, Minneapolis, MN\\
			Emails: \{demir035, yaman013, dowdl016, moell018, lvizioli, yaco0006, strupp, caolman, ugurb001, akcakaya\}@umn.edu}
	}

\maketitle

\begin{abstract}

Functional MRI (fMRI) is commonly used for interpreting neural activities across the brain. Numerous accelerated fMRI techniques aim to provide improved spatiotemporal resolutions. Among these, simultaneous multi-slice (SMS) imaging has emerged as a powerful strategy, becoming a part of large-scale studies, such as the Human Connectome Project. However, when SMS imaging is combined with in-plane acceleration for higher acceleration rates, conventional SMS reconstruction methods may suffer from noise amplification and other artifacts. Recently, deep learning (DL) techniques have gained interest for improving MRI reconstruction. However, these methods are typically trained in a supervised manner that necessitates fully-sampled reference data, which is not feasible in highly-accelerated fMRI acquisitions. Self-supervised learning that does not require fully-sampled data has recently been proposed and has shown similar performance to supervised learning. However, it has only been applied for in-plane acceleration. Furthermore the effect of DL reconstruction on subsequent fMRI analysis remains unclear. In this work, we extend self-supervised DL reconstruction to SMS imaging. Our results on prospectively 10-fold accelerated 7T fMRI data show that self-supervised DL reduces reconstruction noise and suppresses residual artifacts. Subsequent fMRI analysis remains unaltered by DL processing, while the improved temporal signal-to-noise ratio produces higher coherence estimates between task runs.

\end{abstract}


\section{Introduction}

Functional magnetic resonance imaging (fMRI) has played a critical role in improving our understanding of the human brain. While advanced acquisition and reconstruction methods are frequently used in fMRI \cite{uugurbil2013pushing}, further improvements for rapid whole-brain coverage with higher spatiotemporal resolution may greatly benefit neuroscientific studies.

Among numerous accelerated fMRI methods, simultaneous multi-slice (SMS) imaging is commonly used for its ability to provide rapid whole brain coverage at high resolutions \cite{moeller2010multiband,setsompop2012blipped}, even becoming a standard part of large-scale studies, such as the Human Connectome Project (HCP) \cite{van2013wu}. SMS imaging is generally combined with in-plane acceleration to achieve higher acceleration rates while maintaining a reasonable echo time at ultrahigh field strengths \cite{uugurbil2013pushing}. However, at high acceleration rates, conventional linear SMS and parallel imaging reconstruction methods  \cite{setsompop2012blipped,cauley2014interslice,moeller2010multiband} suffer from spatially varying noise amplification \cite{hamilton2017recent,barth2016simultaneous}, and may suffer from both inter-slice and intra-slice residual aliasing artifacts  \cite{todd2016evaluation}. Thus, improvements in image reconstruction may be beneficial, especially in deeper brain areas where higher acceleration rates lead to noise amplification.

Concurrently, outside of fMRI acquisitions, deep learning (DL) methods have recently gained substantial interest as a means to improve accelerated MRI \cite{knoll2020deep,hammernik2018learning,schlemper2017deep,aggarwal2018modl}. Among various approaches, physics-guided DL methods have shown promising results at higher acceleration rates with improved reconstruction quality where conventional methods struggle to maintain high-quality reconstruction \cite{hammernik2018learning,schlemper2017deep,aggarwal2018modl,hosseini2020dense}. Training in deep learning reconstruction is generally performed in a supervised-manner which requires fully-sampled data as reference \cite{knoll2020deep,hammernik2018learning,schlemper2017deep,aggarwal2018modl}. However, the lack of reference training data in highly-accelerated fMRI
acquisitions inhibits such conventional supervised training. Nevertheless, a recent work on self-supervised learning, which trains physics-guided DL reconstruction using only undersampled data \cite{yaman2020self,yaman2020selfisbi}, has shown similar performance to supervised training. Furthermore, the effect of DL reconstruction in general, and self-supervised learning in particular, on subsequent fMRI analysis is not clear, as such analyses capitalize on subtle changes in image characteristics.

In this work, we extend a recent self-supervised DL \cite{yaman2020self} method to SMS reconstruction on HCP-style retinotopic mapping acquisition. The proposed DL technique is compared to conventional SMS reconstruction technique using 10-fold accelerated 7T fMRI data. 
Our results show that the proposed self-supervised physics-guided DL reconstruction reduces residual artifacts and substantially improves temporal SNR, without altering the subsequent fMRI analysis. 

\section{Methods}\label{ssdu}

\subsection{Self-supervised DL Reconstruction for SMS fMRI}

Let $\bf{y}_{SMS}^\Omega \in {\mathbb C}^P$ be the acquired SMS k-space with $\Omega$ as the in-plane undersampling pattern, and ${\bf x}^{i} \in {\mathbb C}^{M \times N}$ be the image corresponding to the $i^{th}$ simultaneously excited slice. The forward model for SMS acquisition is given as: 
\begin{equation}
    \mathbf{y}_{SMS}^{\Omega} =  \sum _{i=1}^{S}\mathbf{E}_{i}^{\Omega}\mathbf{x}_{i} + {\bf n},
    \label{eq1}
\end{equation}
where $\bf{E}_i^{\Omega}: {\mathbb C}^{M \times N} \to  {\mathbb C}^P$ is the multi-coil encoding operator of the $i^{th}$ slice, $S$ is number of simultaneously excited slices and ${\bf n}$ is measurement noise.
Letting ${\bf x}_{SMS}$ be the concatenation of the individual slices $\{{\bf x}_i\}_{i=1}^S$ along the readout direction \cite{moeller2010multiband,demirel2021improved} and $\mathbf{E}_{SMS}^{\Omega} = [\mathbf{E}_{1}^{\Omega} \dots \mathbf{E}_{S}^{\Omega}]$ yields a compact form of Eq. \ref{eq1} as
\begin{equation}
    \mathbf{y}_{SMS}^{\Omega} =  \mathbf{E}_{SMS}^{\Omega}\mathbf{x}_{SMS} + {\bf n}.
    \label{eq2}
\end{equation}
The inverse problem for image reconstruction can be formulated using regularized least squares:
\begin{equation} \label{Eq:recons1}
    \arg \min_{{\bf x}_{SMS}} ||\mathbf{y}_{SMS}^{\Omega} -  \mathbf{E}_{SMS}^{\Omega}\mathbf{x}_{SMS}||_2^2 + {\cal R}(\mathbf{x}_{SMS}),
\end{equation}
where the quadratic term enforces data consistency (DC) with the acquired k-space measurements and $\cal{R}(\cdot)$ is a regularizer. Iterative strategies that decouple the DC term and the regularizer term into a series of sub-problems are commonly used to solve such objective functions \cite{fessler2020optimization}, including variable splitting with quadratic penalty \cite{aggarwal2018modl,yaman2020self}: 
\begin{align}
\mathbf{z}_{SMS}^{(l-1)} &= \arg \min_{\mathbf{z}_{SMS}} \mu\|  \mathbf{x}_{SMS}^{(l-1)} - \mathbf{z}_{SMS} \|_2^2 + {\cal{R}}(\mathbf{z}_{SMS}), \label{Req} \\
    \mathbf{x}_{SMS}^{(l)} &= \arg \min_{\mathbf{x}_{SMS}} \|\mathbf{y}^{\Omega}_{SMS}-\mathbf{E}_{SMS}^{\Omega}\mathbf{x}_{SMS}\|^2_2 \nonumber \\ &\quad\quad\quad\quad\quad+ \mu \| \mathbf{x}_{SMS} - \mathbf{z}_{SMS}^{(l-1)} \|_2^2, \label{DCeq}
\end{align}
where $\mathbf{x}_{SMS}^{(l)}$ is the reconstructed slices at iteration $l$, while $\mathbf{z}_{SMS}^{(l)}$ is an auxiliary variable and $\mu$ is the penalty parameter. 
In physics-guided deep learning reconstruction, this iterative algorithm that alternates between Eq. \ref{Req} and \ref{DCeq} is unrolled for a fixed number of iterations \cite{knoll2020deep}. Though the DC sub-problem (\ref{DCeq}) has a closed-form solution, it is solved using iterative linear methods, such as the conjugate gradient \cite{aggarwal2018modl} due to the difficulty of matrix inversion at these high dimensions. The sub-problem (\ref{Req}) is solved implicitly using neural networks \cite{knoll2020deep}. Overall, the output of the unrolled network, parametrized by ${\bm \theta}$, can be written as $f(\mathbf{y}_{SMS}^{\Omega},\mathbf{E}_{SMS}^{\Omega};\bm{\theta})$.

\begin{figure}[!t]
 \begin{center}
          \includegraphics[trim={0 0 0.7cm 0.4cm},clip, width=\columnwidth]{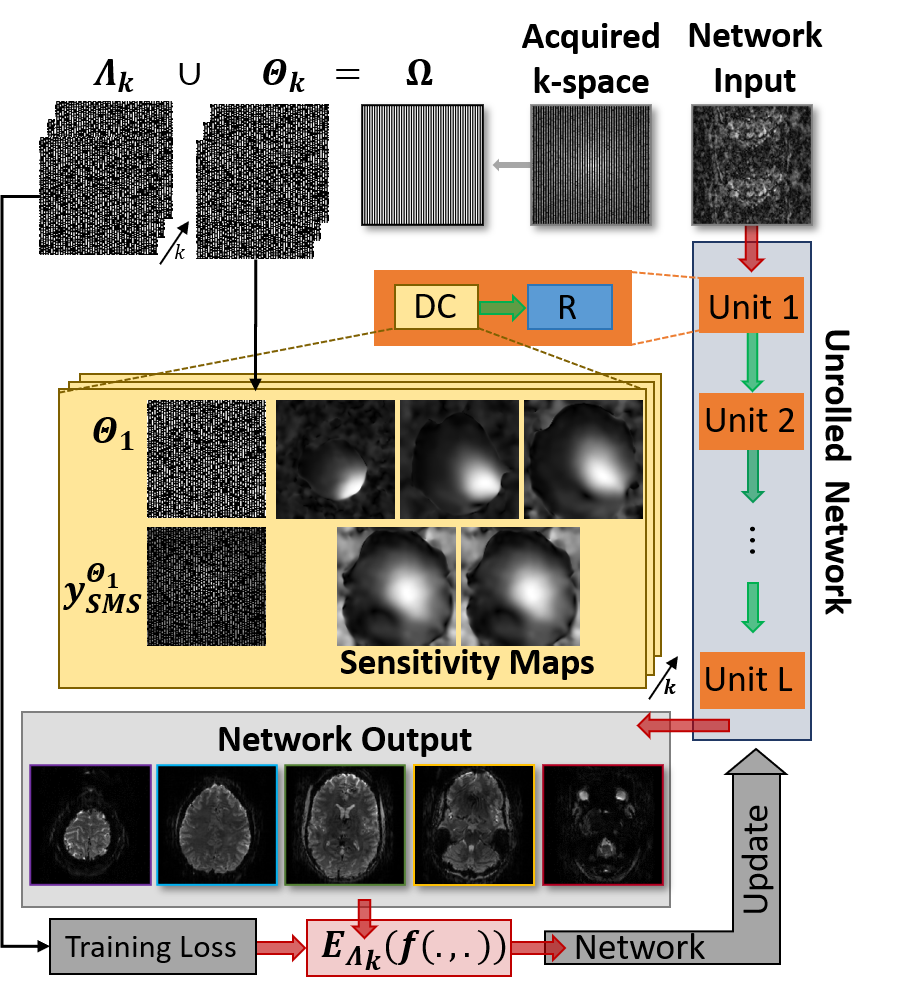}
     \end{center}
      \vspace{-.35cm}
  	\caption{A schematic of our self-supervised learning approach that does not require fully-sampled data. The iterative algorithm using Equation \ref{Req} and \ref{DCeq} is unrolled for a fixed number of iterations. Each iteration consists of a data consistency unit that enforces fidelity with the acquired k-space measurements,  followed by a regularizer unit that is implemented via a ResNet. The self-supervised training divides the acquired samples into multiple disjoint sets ($\Omega = \Theta_k \cup \Lambda_k$), where $\{\Theta_k\}$ are used for data-consistency in the unrolled network, and $\{\Lambda_k\}$ are used for defining training loss. This self-supervised strategy allows training of DL reconstruction for highly-accelerated fMRI, where full-sampled reference data acquisition is not feasible.}
  	\label{fig:1}
  	\vspace{-.35cm}
\end{figure}

Physics-guided neural networks are generally trained in a supervised manner using reference data. However, this strategy is not feasible in high-resolution fMRI, as no such reference data is available. A recently proposed technique, Self-Supervised learning via Data Undersampling (SSDU), tackles the problem of training such neural networks without fully-sampled data  \cite{yaman2020self,yaman2020selfisbi}. It splits the acquired k-space locations, $\Omega$, into two disjoint sets, $\Theta$ and $\Omega$. $\Theta$ is used in the DC units of the unrolled network and $\Lambda$ is used to define the k-space loss.   where one is used to enforce data consistency in the unrolled network, while the other is used to calculate the training loss. SSDU was shown to have similar performance to supervised training \cite{yaman2020self}. Consecutively, a multi-mask version of SSDU was proposed for further improving reconstruction quality at higher acceleration rates \cite{yaman2020multi,yaman2020multiisbi}. Instead of dividing $\Omega$ into only two sets, multi-mask SSDU divides the acquired points into multiple disjoint sets ($\Omega = \Theta_k \cup \Lambda_k$, $k \in {1,\cdots,K}$) as an alternative approach for data augmentation \cite{yaman2020multi}.

In this work, we adapt multi-mask SSDU to SMS imaging for highly-accelerated fMRI, with the following loss function:
\begin{equation}
    \min_{\bm{\theta}} \frac{1}{N \cdot K}\sum_{n=1}^{N}\sum_{k=1}^{K}\mathcal{L}(\mathbf{y}_{SMS}^{\Lambda_k,n},\mathbf{E}_{SMS}^{\Lambda_k,n}(f(\mathbf{y}_{SMS}^{\Theta_k,n},\mathbf{E}_{SMS}^{\Theta_k,n};\bm{\theta}))),
\end{equation}
where $N$ is the number of training data in the database and $\mathcal{L}(\cdot,\cdot)$ is a k-space loss. The $k^{th}$ training and loss masks used on the $n^{th}$ training data sample are depicted with $\Lambda_k,n$ and $\Theta_k,n$, respectively. A schematic of this implementation is shown in Figure \ref{fig:1}.

\subsection{Imaging Experiments} 

\begin{figure*}[!t]
 \begin{center}
          \includegraphics[trim={0 0 0 0},clip, width=6.7 in]{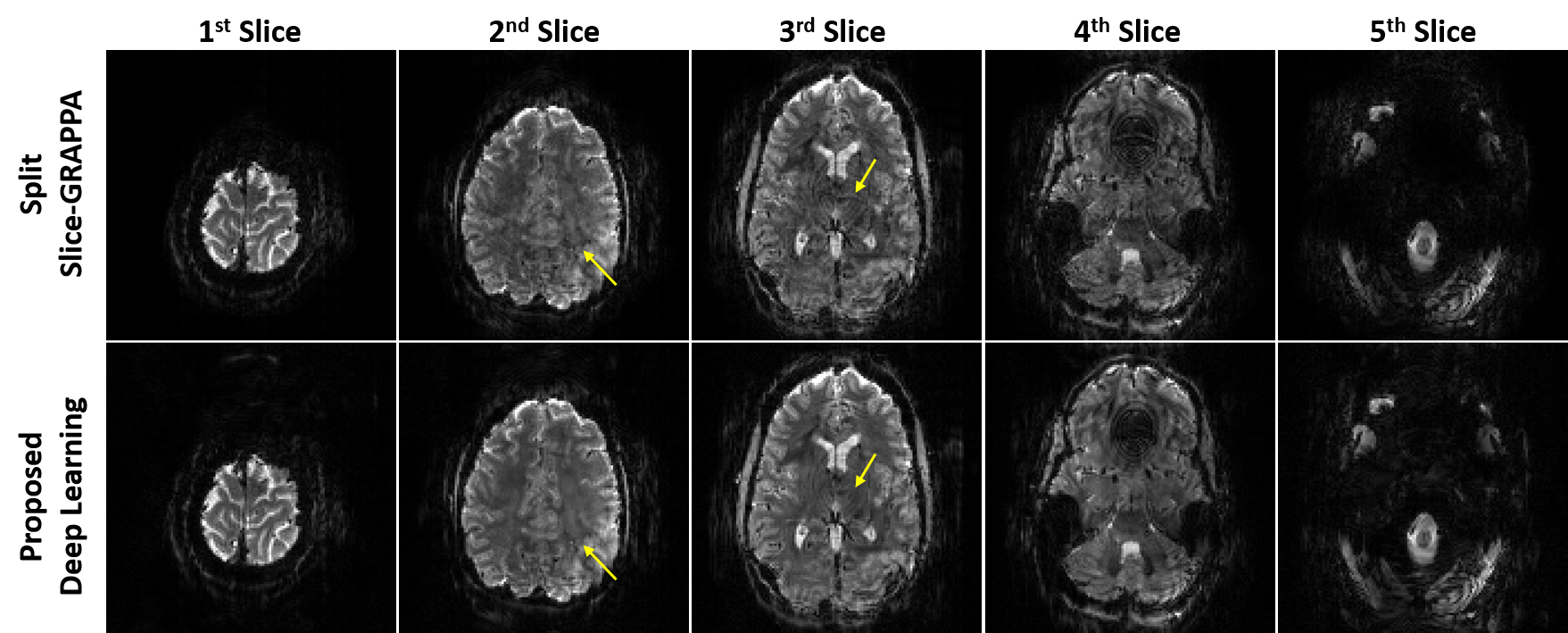}
     \end{center}
      \vspace{-.35cm}
  	\caption{Representative slices from the HCP-style acquisition reconstructed with the conventional split slice-GRAPPA and the proposed self-supervised physics-guided DL reconstruction at 5-fold SMS and 2-fold in-plane acceleration. Proposed deep learning reconstruction reduces noise and suppresses the residual artifacts (shown by yellow arrows) compared to conventional split slice-GRAPPA reconstruction.}
  	\label{fig:2}
  	\vspace{-.35cm}
\end{figure*}

Imaging was performed on a 7T Siemens Magnex Scientific (Siemens Healthineers, Erlangen, Germany) system using a 32-channel receiver head coil-array in 8 subjects. HCP-retinotopy protocol was used \cite{benson2018human}. The study was approved by our institutional review board. Written informed consents were obtained before each examination. 

Six 5-minute retinotopic mapping runs were acquired for each subject, with the experimental protocol detailed in \cite{benson2018human}. 
A 5-fold SMS and 2-fold in-plane acceleration was utilized at 1.6mm isotropic resolution with 1s TR and whole-brain coverage, similar to the 7T HCP fMRI protocol \cite{uugurbil2013pushing}. In this work, we focus on the first 2 experimental runs, entailing a rotating wedge (RETCCW, RETCW) retinotopic mapping paradigm. The RETCCWs were collected with an anterior to posterior phase encoding direction, whereas RETCWs were collected with a posterior to anterior phase encoding direction. In both acquisitions, wedges that complete a rotation every 32 seconds were seen by the subjects. These data were used to perform retinotopic mapping of the visual cortex and further detailed analysis on the full dataset can be found in \cite{benson2018human}.

\begin{figure}[!b]
 \begin{center}
          \includegraphics[trim={0 0 0 0},clip, width=3.0 in]{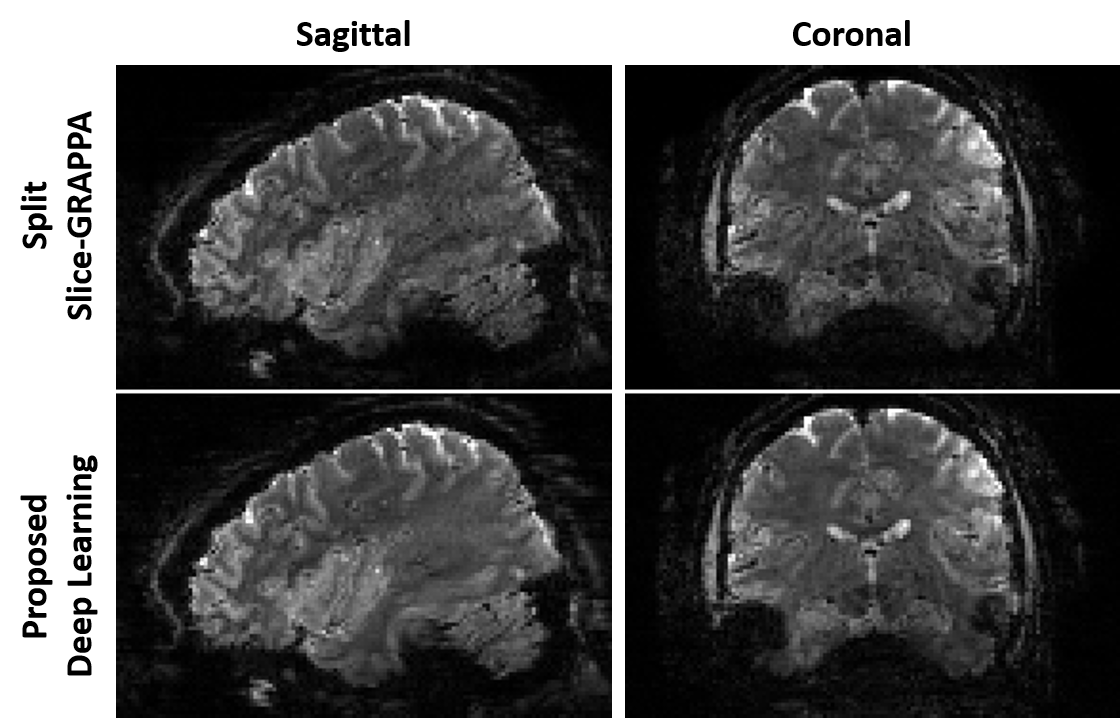}
     \end{center}
      \vspace{-.35cm}
  	\caption{Coronal and sagittal slices using conventional split slice-GRAPPA and the proposed self-supervised DL reconstructions. Deep learning visibly improves image quality compared to split slice-GRAPPA. Substantial noise reduction is observed in mid-brain with the proposed DL reconstruction.}
  	\label{fig:3}
  	\vspace{-.35cm}
\end{figure}

\subsection{Reconstruction Implementation Details}

Acquired raw k-space data were exported from the scanner, and all training and reconstructions were performed off-line. Physics-guided DL was trained using multi-mask SSDU \cite{yaman2020multi,yaman2020selfisbi,yaman2020multiisbi} with $K$= 6 masks. The network was unrolled for 10 iterations where each iteration involved a DC unit and a regularizer. Conjugate gradient was used in the DC units \cite{aggarwal2018modl}, whereas the regularization was implicitly performed by a convolutional neural network that utilizes a ResNet structure \cite{timofte2017ntire}. Sensitivity maps were generated from the low-resolution calibration scans using ESPIRiT \cite{uecker2014espirit}. Adam optimizer with an $\ell_1$-$\ell_2$ loss in k-space \cite{yaman2020self} and a learning rate of $3\cdot10^{-4}$ over 100 epochs were used to train the network. Training was performed using TensorFlow in Python. SMS = 5 slices were concatenated along the readout direction, as detailed in Section \ref{ssdu}, and the proper FOV shifts were applied before and after the ResNet to avoid boundary artifacts. 17 slab groups from 4 subjects with only one time-frame per subject were used for training. Thus, no temporal information was shared. 4 different subjects unseen by the network were used for testing. The whole volume and the whole time series of the test subjects were processed. Comparisons were made to an implementation of split slice-GRAPPA reconstruction \cite{cauley2014interslice}, which is the conventionally used reconstruction method for such HCP-type acquisitions.

\begin{figure}[!t]
 \begin{center}
          \includegraphics[trim={0.2cm 0.2cm 0 0},clip, width=\columnwidth]{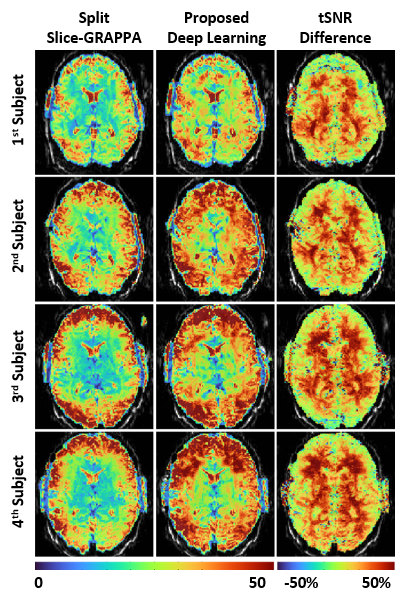}
     \end{center}
      \vspace{-.45cm}
  	\caption{Temporal SNR (tSNR) maps of axial views from 4 test subjects using split slice-GRAPPA and the proposed self-supervised DL reconstructions. An overall tSNR gain is observed in self-supervised DL reconstruction compared to split slice-GRAPPA in all subjects. tSNR difference calculated in units of \% change shows that proposed DL reconstruction improved upon split slice-GRAPPA with up to 50\% gain, particularly in deep brain areas.}
  	\label{fig:4}
  	\vspace{-.35cm}
\end{figure}

\subsection{Functional Processing and Retinotopic Analyses}

Functional image preprocessing was performed using \texttt{afni\_poc.py}, a tool available with AFNI \cite{cox1996afni}. Images were corrected first for distortion arising from differences in the phase encoding direction, and subsequently for motion using rigid-body alignment (6 degrees of freedom). The data were then scaled to have a mean of 100 for each voxel and imported into MATLAB. The clockwise wedge data were time-reversed and temporally shifted to align the hemodynamic responses for identical wedge positions. Next, 9 nuisance regressors were projected out from each dataset. These were polynomials up to order 3 and the 6 motion estimates from the motion-correction step.  The mean of the two datasets was calculated and FFT was performed to determine the amplitude and phase for the 32 second rotations (0.3125Hz). Coherence between the two runs were calculated using the MATLAB tool \texttt{mscohere} to threshold the data. The phase maps were compared for each reconstruction method using a coherence threshold of 0.55. Additionally, the mean of the signal after processing divided by the standard deviation of the residual following the removal of nuisance regressors was used to calculate the temporal signal-to-noise ratio (tSNR).

\begin{figure}[!]
 \begin{center}
          \includegraphics[trim={0 0 0 0.2cm},clip, width=2.98 in]{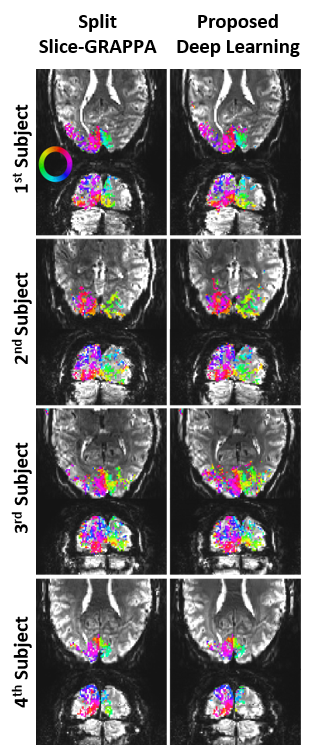}
     \end{center}
      \vspace{-.35cm}
  	\caption{Phase maps for each reconstruction type with a coherence threshold of 0.55, Color indicates polar angle, in radians (legend on the top left image). Right side of the brain is on the right. The phase estimates show strong agreement between split slice-GRAPPA and the proposed self-supervised DL reconstructions. Additionally, the DL reconstruction has more voxels above the threshold, which may be correlated with its improved tSNR.
  	 }
  	\label{fig:5}
  	\vspace{-.4cm}
\end{figure}

\section{RESULTS}

Figure \ref{fig:2} depicts representative reconstructions from the proposed self-supervised physics-guided DL and conventional split slice-GRAPPA  for the 10-fold accelerated fMRI acquisition. Five axial slices from one slab group are shown, where the proposed DL reconstruction visibly reduces noise, while suppressing residual artifacts. Particularly, significant improvement is observed in the thalamus region (3rd slice, yellow arrows) with the proposed DL method. Coronal and sagittal views of the whole brain are shown in Figure \ref{fig:3}, where visible noise reduction in mid-brain is observed with the proposed DL strategy. 

Axial tSNR maps of 4 subjects reconstructed with split slice-GRAPPA and self-supervised DL are shown in Figure \ref{fig:4}. A substantial overall tSNR improvement is observed with the proposed DL reconstruction in white matter, and in the thalamus along with its nearby areas. The last column depicts tSNR differences calculated in units of $\%$ change, where the proposed method shows approximately $50\%$ higher tSNR compared to split slice-GRAPPA, especially in mid-brain areas.

Figure \ref{fig:5} shows the estimated phase sensitivity in the cortical surface based on a threshold of 0.55 coherence. The two runs of the wedge task show strong agreement between split slice-GRAPPA and physics-guided DL reconstruction. The voxels that respond best for a stimulus in one area of the visual field show the same responses and the phase estimates are indistinguishable between split slice-GRAPPA and DL reconstructions. However, the DL reconstruction appears to have more voxels above the threshold, possibly due to the tSNR improvement. Additionally, high noise levels associated with high acceleration rates in areas further from receiver coils are mitigated with the proposed method. These tSNR gains are most pronounced in areas such as the thalamus and the cerebellum, which are typically associated with poor signal quality.

\section{Conclusion}

In this work, we proposed a self-supervised physics-guided deep learning reconstruction for highly-accelerated HCP-style fMRI acquisitions. For prospectively 5-fold SMS and 2-fold in-plane accelerated fMRI data, the self-supervised DL reconstruction reduced residual artifacts and improved tSNR compared to the conventional split slice-GRAPPA reconstruction approach. Subsequent fMRI analysis showed that the DL reconstruction did not alter the temporal affects or functional spatial precision, which has been a concern for more traditional regularized reconstruction methods like compressed sensing \cite{chiew2018recovering,fang2016high}. Thus, the proposed self-supervised learning approach allows the application of DL methods to highly-accelerated fMRI, and further investigations at higher acceleration rates are warranted to harness the full potential of the proposed DL reconstruction.

\section*{ACKNOWLEDGMENTS}
Grant support: NIH, Grant numbers: R01HL153146, U01EB025144, P41EB027061, P30NS076408; NSF, Grant number: CAREER CCF-1651825.

\bibliographystyle{IEEEbib}
\bibliography{reference}

\end{document}